\documentclass[11pt]{amsart}
\DeclareMathAlphabet{\EuFrak}{U}{euf}{m}{n}
\DeclareMathAlphabet{\EuScript}{U}{eus}{m}{n}
\usepackage{amsmath}
\usepackage{amsfonts,amssymb,latexsym}
\usepackage{mathrsfs}
\usepackage{amscd}
\usepackage{amsthm}
\newtheorem{theorem}{\rmfamily\bfseries{Theorem}}[section]

\newtheorem{lemma}[theorem]{\rmfamily\bfseries{Lemma}} 
\newtheorem{proposition}[theorem]{\rmfamily\bfseries{Proposition}}

\newtheorem{definition}[theorem]{\rmfamily\bfseries{Definition}}

\theoremstyle{remark}

\newtheorem{remark}{Remark}

\usepackage{graphics}
\numberwithin{equation}{section}
\def\Bbb{\mathbb}
\newcommand{\mona}{\mbox{\LARGE\itshape a}}
\def\logo{\raisebox{-10.5\p@}{\hb@xt@85\p@{\includegraphics{gft.eps}\hfil}}}

\def\un{1\kern-3pt \rm I}

\newcommand{\oN}{{\mathbb N}}
\newcommand{\oR}{{\mathbb R}}

\newcommand{\oC}{{\mathbb C}}

\newcommand{\supp}{\mathrm{supp}}
   
\begin{document}
{\hfill
\parbox{50mm}{{\sf CEFT-SFM-DHTF06/3}\\{\sf Revised version}} \vspace{8mm}}

\title[Reeh-Schlieder Theorem]
      {{Reeh-Schlieder Theorem for \\ Ultrahyperfunctional Wightman Theory}}

\author{Daniel H.T. Franco}
\address{Universidade Federal de Vi\c cosa \\
        Departamento de F\'\i sica, Avenida Peter Henry Rolfs s/n \\  
        Campus Universit\'ario, Vi\c cosa, MG, Brasil, CEP: 36570-000.}
\email{dhtf@terra.com.br}

\date{\today}
\thanks{This work is supported by the Funda\c c\~ao de Amparo \`a Pesquisa do
        Estado de Minas Gerais (FAPEMIG) agency, grant CEX00012/07. Also at
        Centro de Estudos de F\'\i sica Te\'orica, Setor de F\'\i sica--Matem\'atica,
        Belo Horizonte, MG, Brasil.}

\begin{abstract}
It will be shown that the Reeh-Schlieder property holds for states of quantum
fields for ultrahyperfunctional Wightman theory. As by product, it is shown
that the Reeh-Schlieder property also holds for states of quantum
fields on a non-commutative Minkowski space in the setting ultrahyperfunctional. 
\end{abstract}

\maketitle

\section{Introduction}
\label{Section1}
In recent years a considerable effort has been made to clarify the structural
aspects of non-commutative quantum field theories (NCQFT). The first paper on quantum
field theory by exploring the non-commutativity of a space-time manifold was
proposed a long time ago as a generalization of the phase space of quantum mechanics
by Snyder~\cite{Sny}, who used this idea to give a solution for the problem of ultraviolet
divergences which had plagued quantum field theories from very beginning. Since then, due
to the success of the renormalization theory, this subject was abandoned. Only recently the
plan of investigating field theories on non-commutative space-times has been revived. In a
fundamental paper Doplicher-Fredenhagen-Roberts~\cite{DFR} have shown that a model quantum
space-time can be described by a non-commutative algebra whose commutation relations do
imply uncertainty relations motivated by Heisenberg's uncertainty principle and
by Einstein's theory. Later, in a different context, NCQFT appear directly related with
the string theory~\cite{SeiWit}, when was found that a non-commutative Yang-Mills theory
induced by the Moyal product can be seen as a vestige, in the low-energy limit, of open
strings in the presence of a constant magnetic field, $B_{\mu\nu}$
(for a review see~\cite{DouNe,Sza}).

From an axiomatic standpoint, a language has been developed which, in principle,
ought to enable one to extend the Wightman axioms to this context~\cite{AGVM}-\cite{DZH}.
However, the axi\-omatic approach to {\bf local} quantum field theory built up by
Streater-Wightman~\cite{SW}, Jost~\cite{Jost}, Bogoliubov {\em et al.}~\cite{BLOT},
Haag~\cite{Haag} and others turned out to be too narrow for theoretical physicists,
who are interested in handling situations involving a NCQFT. In particular, some very
important evidences to expect that the traditional Wightman axioms must be somewhat modified
for the setting of NCQFT are:

\begin{itemize}

\item NCQFT incorporate {\bf nonlocal} effects, but in a controllable way.
This is reminiscent of its stringy origin where the gravitational sector was decoupled
but still left some traces through the non-commutativity.

\item The existence of hard infrared singularities in the non-planar sector
of the theory can destroy the {\bf tempered} nature of the Wightman functions.

\item The commutation relations $[x_\mu,x_\nu]=i \theta_{\mu\nu}$
also imply uncertainty relations for space-time coordinates
$\Delta x_\mu \Delta x_\nu \sim \bigl|\theta_{\mu\nu}\bigr|$, indicating that the
notion of space-time point loses its meaning. Space-time points are replaced by cells
of area of size $\bigl|\theta_{\mu\nu}\bigr|$. This suggests the existence of
a finite lower limit to the possible resolution of distance. The {\bf nonlocal}
structure of NCQFT manifests itself in a indeterminacy of the interaction regions,
which spread over a space-time domain whose size is determined by
the existence of a {\bf fundamental length} $\ell$ related to the scale of nonlocality
$\ell \sim \sqrt{\bigl|\theta_{\mu\nu}\bigr|}$. 

\end{itemize}

In Ref.~\cite{DZH} has been suggested that tempered ultrahyperfunctions
corresponding to tubular radial domains are well adapted for their use in
the axiomatic description of NCQFT. The space of tempered ultrahyperfunctions has the
advantage of being representable by means of holomorphic functions. It is the dual space
of the space of entire functions rapidly decreasing in any horizontal strip
and generalizes the notion of hyperfunctions on $\oR^n$, but {\em can not} be localized
as hyperfunctions. In the framework of this approach, fundamental results, as the
CPT and Spin-Statistics theorems, the Borchers class of a non-commutative field
and the Reconstruction theorem, were proven~\cite{DZH}.

In this article we prove that the Reeh-Schlieder-type property~\cite{ReehSchlieder}
holds for states of quantum fields for ultrahyperfunctional Wightman theory. Then,
as by product, it is shown that the Reeh-Schlieder property also holds for states of
quantum fields on a non-commutative Minkowski space in the setting ultrahyperfunctional. 
According to the standard arguments, the Reeh-Schlieder property concerns with the cyclicity
and separability of the vacuum sector in the context of local quantum field theories in
Minkowski space-time. However, it holds equally well for a quantum field theory on curved
space-times~\cite{Verch}-\cite{StV}, as well as for thermal states~\cite{Jakel}
as a direct consequence of locality, additivity and the relativistic KMS condition.
Once one has the concept of fundamental length incorporated in NCQFT, a natural
problem is to recognize whether the Reeh-Schlieder property can also be
established for a non-commutative quantum field theory. We show that this is feasible
since a crucial mathematical tool leading to the Reeh-Schlieder property
in the case of NCQFT is a tempered ultrahyperfunction version of Edge of the
Wedge theorem~\cite{Daniel1}.

We outline the content of this contribution as follows.
In Section \ref{Section2}, for the convenience of the reader, we shall present
briefly some definitions and basic properties of the
tempered ultrahyperfunction space of Sebasti\~ao e Silva~\cite{Tiao1,Tiao2}
and Hasumi~\cite{Hasumi} (we indicate the Refs.~\cite{Tiao1}-\cite{Daniel1} for more
details). Section \ref{Sect3} contains some needed results concerning with the
proof of the Reeh-Schlieder theorem for ultrahyperfunctional Wightman Theory. In Section
\ref{Section4}, we give Reeh-Schlieder theorem for ultrahyperfunctional Wightman Theory
and for NCQFT. Throughout the paper we assume only the case of space-space non-commutativity,
{\em i.e.}, $\theta_{0i}=0$, with $i=1,2,3$. It is well known that if there is space-time
non-commutativity, the resulting theory violates the causality and unitarity~\cite{SST,Gomis}.
We consider for simplicity a theory with only one basic field, a neutral scalar field.
Section \ref{Section5} contains the final considerations.

\section{Tempered Ultrahyperfunctions: Some Basic Properties}
\label{Section2}
Tempered ultrahyperfunctions were introduced in papers of Sebasti\~ao e
Silva \cite{Tiao1,Tiao2} and Hasumi~\cite{Hasumi} (orginally called {\em tempered
ultradistributions}) as the strong dual of the space of test functions
of rapidly decreasing entire functions in any horizontal strip.
While Sebasti\~ao e Silva~\cite{Tiao1} used extension
procedures for the Fourier transform combined with holomorphic representations
and considered the 1-dimensional case, Hasumi~\cite{Hasumi} used
duality arguments in order to extend the notion
of tempered ultrahyperfunctions for the case of $n$ dimensions
(see also~\cite[Section 11]{Tiao2}). In a brief tour, Marimoto~\cite{Mari1,Mari2}
gave some more precise informations concerning the work of Hasumi.
More recently, the relation between the tempered ultrahyperfuntions and
Schwartz distributions and some major results, as the kernel theorem and the
Fourier-Laplace transform have been established by Br\"uning and Nagamachi
in~\cite{BruNa1}. Earlier, some precisions on the Fourier-Laplace transform
theorem for tempered ultrahyperfunctions were given by Carmichael~\cite{Carmi2}
(see also~\cite{DanHenri,Daniel1}), by considering the theorem in its simplest form,
{\em i.e.}, the equivalence between support properties of a distribution
in a closed convex cone and the holomorphy of its Fourier-Laplace transform
in a suitable tube with conical basis. In this more general setting, which
includes the results of Sebasti\~ao e Silva and Hasumi as special cases, Carmichael
obtained new representations of tempered ultrahyperfunctions which were not considered
by Sebasti\~ao e Silva~\cite{Tiao1,Tiao2} or Hasumi~\cite{Hasumi}.
In this section, we include the definitions and basic pro\-per\-ties of the
tempered ultrahyperfunction space which are the most important in
applications to quantum field theory.

Next, we shall introduce briefly here some definitions and basic properties of the
tempered ultrahyperfunction space of Sebasti\~ao e Silva~\cite{Tiao1,Tiao2}
and Hasumi~\cite{Hasumi} (we indicate the Refs. for more details).
To begin with, we introduce the following multi-index notation. Let
$\oR^n$ (resp. $\oC^n$) be the real (resp. complex) $n$-space whose generic points
are denoted by $x=(x_1,\ldots,x_n)$ (resp. $z=(z_1,\ldots,z_n)$), such that
$x+y=(x_1+y_1,\ldots,x_n+y_n)$, $\lambda x=(\lambda x_1,\ldots,\lambda x_n)$,
$x \geq 0$ means $x_1 \geq 0,\ldots,x_n \geq 0$, $\langle x,y \rangle=x_1y_1+\cdots+x_ny_n$
and $|x|=|x_1|+\cdots+|x_n|$. Moreover, we define
$\alpha=(\alpha_1,\ldots,\alpha_n) \in \oN^n_o$, where $\oN_o$ is the set
of non-negative integers, such that the length of $\alpha$ is the corresponding
$\ell^1$-norm $|\alpha|=\alpha_1+\cdots +\alpha_n$, $\alpha+\beta$ denotes
$(\alpha_1+\beta_1,\ldots,\alpha_n+\beta_n)$, $\alpha \geq \beta$ means
$(\alpha_1 \geq \beta_1,\ldots,\alpha_n \geq \beta_n)$, $\alpha!=
\alpha_1! \cdots \alpha_n!$, $x^\alpha=x_1^{\alpha_1}\ldots x_n^{\alpha_n}$,
and
\[
D^\alpha \varphi(x)=\frac{\partial^{|\alpha|}\varphi(x_1,\ldots,x_n)}
{\partial x_1^{\alpha_1}\partial x_2^{\alpha_1}\ldots\partial x_n^{\alpha_n}}\,\,.
\]
Let $\Omega$ be a set in $\oR^n$. Then we denote by $\Omega^\circ$ the interior
of $\Omega$ and by $\overline{\Omega}$ the closure of $\Omega$. For $r > 0$, we
denote by $B(x_o;r)=\bigl\{x \in \oR^n \mid |x-x_o| < r\bigr\}$ a open ball
and by $B[x_o;r]=\bigl\{x \in \oR^n \mid |x-x_o| \leq r\bigr\}$ a closed ball,
with center at point $x_o$ and of radius $r=(r_1,\ldots,r_n)$, respectively.

We consider two $n$-dimensional spaces -- $x$-space and $\xi$-space -- with the
Fourier transform defined
\[
\widehat{f}(\xi)={\mathscr F}[f(x)](\xi)=
\int_{\oR^n} f(x)e^{i \langle \xi,x \rangle} d^nx\,\,,
\]
while the Fourier inversion formula is
\[
f(x)={\mathscr F}^{-1}[\widehat{f}(\xi)](x)= \frac{1}{(2\pi)^n}
\int_{\oR^n} \widehat{f}(\xi)e^{-i \langle \xi,x \rangle} d^n\xi\,\,.
\]
The variable $\xi$ will always be taken real while $x$ will also be
complexified -- when it is complex, it will be noted $z=x+iy$. The
above formulas, in which we employ the symbolic ``function notation,''
are to be understood in the sense of distribution theory.

We shall consider the function
\[
h_{K}(\xi)=\sup_{x \in K} \langle \xi,x \rangle\,\,,
\quad \xi \in \oR^n\,\,,
\]
where $K$ is a compact set in $\oR^n$. One calls $h_{K}(\xi)$ the {\it supporting
function} of $K$. We note that $h_{K}(\xi) < \infty$ for every $\xi \in \oR^n$ since
$K$ is bounded. For sets $K=\bigl[-k,k\bigr]^n$, $0 < k < \infty$, the supporting
function $h_{K}(\xi)$ can be easily determined:
\[
h_{K}(\xi)=\sup_{x \in K} \langle \xi,x \rangle =
k|\xi|\,\,,\quad \xi \in \oR^n\,\,,\quad |\xi|=\sum_{i=1}^n|\xi_i|\,\,.
\]

Let $K$ be a convex compact subset of $\oR^n$,
then $H_b(\oR^n;K)$ ($b$ stands for bounded) defines the space of all
functions $\in C^\infty(\oR^n)$ such that $e^{h_K(\xi)}D^\alpha\!f(\xi)$
is bounded in $\oR^n$ for any multi-index $\alpha$. One defines in
$H_b(\oR^n;K)$ seminorms
\begin{equation}
\|\varphi\|_{K,N}=\sup_{x \in \oR^n; \alpha \leq N}
\bigl\{e^{h_K(\xi)}|D^\alpha f(\xi)|\bigr\} < \infty\,\,,
\quad N=0,1,2,\ldots\,\,.
\label{snorma2}
\end{equation}

If $K_1 \subset K_2$ are two compact convex sets, then
$h_{K_1}(\xi) \leq h_{K_2}(\xi)$, and thus the canonical
injection $H_b(\oR^n;K_2) \hookrightarrow H_b(\oR^n;K_1)$
is continuous. Let $O$ be a convex open set of $\oR^n$.
To define the topology of $H(\oR^n;O)$ it suffices to let $K$ range
over an increasing sequence of convex compact subsets $K_1,K_2,\ldots$
contained in $O$ such that for each $i=1,2,\ldots$,
$K_i \subset K_{i+1}^\circ$ and ${O}=\bigcup_{i=1}^\infty K_i$.
Then the space $H(\oR^n;O)$ is the projective limit of the
spaces $H_b(\oR^n;K)$ according to restriction mappings
above, {\em i.e.}
\begin{equation}
H(\oR^n;O)=\underset{K \subset {O}}{\lim {\rm proj}}\,\,
H_b(\oR^n;K)\,\,,
\label{limproj2}
\end{equation} 
where $K$ runs through the convex compact sets contained in $O$.

\begin{theorem}[\cite{Hasumi,Mari1,BruNa1}]
The space ${\mathscr D}({\oR^n})$ of all $C^\infty$-functions
on $\oR^n$ with compact support is dense in $H(\oR^n;K)$ and $H(\oR^n;O)$.
Moreover, the space $H(\oR^n;\oR^n)$ is dense in $H(\oR^n;O)$ and
$H(\oR^m;\oR^m) \otimes H(\oR^n;\oR^n)$ is dense in $H(\oR^{m+n};\oR^{m+n})$.
\label{theoINJ}
\end{theorem}

From Theorem \ref{theoINJ} we have the following injections~\cite{Mari1}:
\[
H^\prime(\oR^n;K) \hookrightarrow H^\prime(\oR^n;\oR^n)
\hookrightarrow {\mathscr D}^\prime(\oR^n)\,\,,
\]
and
\[
H^\prime(\oR^n;O) \hookrightarrow H^\prime(\oR^n;\oR^n)
\hookrightarrow {\mathscr D}^\prime(\oR^n)\,\,.
\]

A distribution $V \in H^\prime(\oR^n;O)$ may be expressed as a finite
order deri\-va\-ti\-ve of a continuous function of exponential growth
\[
V=D^\gamma_\xi[e^{h_K(\xi)}g(\xi)]\,\,,
\]
where $g(\xi)$ is a bounded continuous function. For $V \in
H^\prime(\oR^n;O)$ the follo\-wing result is known:

\begin{lemma}[\cite{Mari1}]
A distribution $V \in {\mathscr D}^\prime(\oR^n)$ belongs to $H^\prime(\oR^n;O)$
if and only if there exists a multi-index $\gamma$, a convex compact set $K \subset O$
and a bounded continuous function $g(\xi)$ such that
\[
V=D^\gamma_\xi[e^{h_K(\xi)}g(\xi)]\,\,.
\]
\label{lemmaMari}
\end{lemma}

In the space $\oC^n$ of $n$ complex variables $z_i=x_i+iy_i$,
$1 \leq i \leq n$, we denote by $T(\Omega)=\oR^n+i\Omega \subset \oC^n$
the tubular set of all points $z$, such that $y_i={\text{Im}}\,z_i$ belongs
to the domain $\Omega$, {\em i.e.}, $\Omega$ is a connected open set in $\oR^n$
called the basis of the tube $T(\Omega)$. Let $K$ be a convex compact
subset of $\oR^n$, then ${\mathfrak H}_b(T(K))$ defines
the space of all continuous functions $\varphi$ on $T(K)$ which are holomorphic
in the interior $T(K^\circ)$ of $T(K)$ such that the estimate
\begin{equation}
|\varphi(z)| \leq {\boldsymbol{\sf M}}_{_{K,N}}(\varphi) (1+|z|)^{-N}
\label{est}
\end{equation}
is valid. The best possible constants in (\ref{est}) are given by a family of
seminorms in ${\mathfrak H}_b(T(K))$
\begin{equation}
\|\varphi\|_{K,N}=\sup_{z \in T(K)}
\bigl\{(1+|z|)^{N}|\varphi(z)|\bigr\} < \infty\,\,,
\quad N=0,1,2,\ldots\,\,.
\label{snorma1}
\end{equation}

If $K_1 \subset K_2$ are two convex compact sets, then
${\mathfrak H}_b(T(K_2)) \hookrightarrow {\mathfrak H}_b(T(K_1)$.
Given that the spaces ${\mathfrak H}_b(T(K_i))$ are Fr\'echet spaces, the space
${\mathfrak H}(T({O}))$ is characterized as a projective limit of Fr\'echet spaces 
\begin{equation}
{\mathfrak H}(T({O}))=\underset{K \subset {O}}{\lim {\rm proj}}\,\,
{\mathfrak H}_b(T(K))\,\,,
\label{limproj1}
\end{equation}
where $K$ runs through the convex compact sets contained in $O$ and
the projective limit is taken following the restriction mappings above.

For any element $U \in {\mathfrak H}^\prime$, its Fourier transform is
defined to be a distribution $V$ of exponential growth, such that the
Parseval-type relation
\begin{equation}
\langle V,\varphi \rangle=\langle U,\psi \rangle\,\,,\quad
\varphi \in H\,\,,\,\,\psi={\mathscr F}[\varphi] \in {\mathfrak H}\,\,,
\label{PRel1}
\end{equation}
holds. In the same way, the inverse Fourier transform of a distribution $V$ of
exponential growth is defined by the relation 
\begin{equation}
\langle U,\psi \rangle=\langle V,\varphi \rangle\,\,,\quad
\psi \in {\mathfrak H}\,\,,\,\,\varphi={\mathscr F}^{-1}[\psi] \in H\,\,.
\label{PRel2}
\end{equation}

\begin{proposition}[\cite{Mari1}]
If $f \in H(\oR^n;O)$, the Fourier transform of $f$ belongs
to the space ${\mathfrak H}(T(O))$, for any open convex
non-empty set $O \subset \oR^n$. By the dual Fourier transform
$H^\prime(\oR^n;O)$ is topologically isomorphic with the space
${\mathfrak H}^\prime(T(-O))$.
\label{Propo1}
\end{proposition}

\begin{definition}
A tempered ultrahyperfunction is a continuous linear functional defined
on the space of test functions ${\mathfrak H}(T(\oR^n))$ of rapidly decreasing
entire functions in any horizontal strip. 
\label{UHF}
\end{definition}

The space of all tempered ultrahyperfunctions is denoted by ${\mathscr U}(\oR^n)$.
As a matter of fact, these objects are equi\-va\-lence classes of holomorphic
functions defined by a certain space of functions which are analytic in the $2^n$
octants in $\oC^n$ and represent a natural generalization of the notion of
hyperfunctions on $\oR^n$, but are {\it non-localizable}. The space
${\mathscr U}(\oR^n)$ is characterized in the following way~\cite{Hasumi}:
Let $\boldsymbol{{\mathscr H}_\omega}$ be the space of all
functions $f(z)$ such that ({\it i}) $f(z)$ is analytic for $\{z \in \oC^n \mid
|{\rm Im}\,z_1| > p, |{\rm Im}\,z_2| > p,\ldots,|{\rm Im}\,z_n| > p\}$,
({\it ii}) $f(z)/z^p$ is bounded continuous  in
$\{z\in \oC^n \mid |{\rm Im}\,z_1| \geqq p,|{\rm Im}\,z_2| \geqq p,
\ldots,|{\rm Im}\,z_n| \geqq p\}$, where $p=0,1,2,\ldots$ depends on $f(z)$
and ({\it iii}) $f(z)$ is bounded by a power of $z$, $|f(z)|\leq
{\boldsymbol{\sf M}}(1+|z|)^N$, where ${\boldsymbol{\sf M}}$ and $N$ depend on
$f(z)$. Define the {\em kernel} of the mapping $f:{\mathfrak H}(T(\oR^n))
\rightarrow \oC$ by $\boldsymbol{\Pi}$, as the set of all $z$-dependent
pseudo-polynomials, $z\in \oC^n$ (a pseudo-polynomial is a
function of $z$ of the form $\sum_s z_j^s G(z_1,...,z_{j-1},z_{j+1},...,z_n)$,
with $G(z_1,...,z_{j-1},z_{j+1},...,z_n) \in \boldsymbol{{\mathscr H}_\omega}$).
Then, $f(z) \in \boldsymbol{{\mathscr H}_\omega}$ belongs to the kernel
$\boldsymbol{\Pi}$ if and only if $\langle f(z),\psi(x) \rangle=0$,
with $\psi(x) \in {\mathfrak H}(T(\oR^n))$ and $x={\rm Re}\,z$.
Consider the quotient space ${\mathscr U}=\boldsymbol{{\mathscr H}_\omega}
/\boldsymbol{\Pi}$. The set ${\mathscr U}$ is the space of tempered
ultrahyperfunctions. Thus, we have the

\begin{theorem}[Hasumi~\cite{Hasumi}, Proposition 5] The space of
tempered ultrahyperfunctions ${\mathscr U}$ is algebraically isomorphic
to the space of generalized functions ${\mathfrak H}^\prime$.
\label{HasumiTheo}
\end{theorem}

\begin{theorem}[Kernel theorem for tempered ultrahyperfunctions~\cite{BruNa1}]
Let $M$ be a separately continuous multilinear functional on $[{\mathfrak H}(T(\oR^4))]^n$.
Then there is a unique functional $F \in {\mathfrak H}^\prime(T(\oR^{4n}))$, for all
$f_i \in {\mathfrak H}(T(\oR^4))$, $i=1,\ldots,n$ such that
$M(f_1,\ldots,f_n)=F(f_1 \otimes \cdots \otimes f_n)$.
\label{KernelTheo} 
\end{theorem}

\begin{theorem}[\cite{Mari1,BruNa1}]
The space ${\mathfrak H}(T(\oR^n))$ is dense in ${\mathfrak H}(T(O))$ and
the space ${\mathfrak H}(T(\oR^{m+n}))$ is dense in ${\mathfrak H}(T(O))$.
\label{theoINJE}
\end{theorem}

\section{Tempered Ultrahyperfunctions Corresponding to a Proper Convex Cone}
\label{Sect3}
In order to prove the theorem Reeh-Schlieder theorem for NCQFT in terms
of tempered ultrahyperfuncions, we shall recall some needed results
taken from Refs.~\cite{DanHenri,Daniel1}.
We now shall define the space of holomorphic functions with which this
paper is concerned. We start by introducing some terminology and simple
facts concerning cones. An open set $C \subset \oR^n$ is called a cone if
$\oR_+ \cdot C \subset C$. A cone $C$ is an open connected cone if $C$ is an
open connected set. Moreover, $C$ is called convex if $C+C \subset C$ and {\it proper}
if it contains no any straight line. A cone $C^\prime$ is called compact in $C$ --
we write $C^\prime \Subset C$ -- if the projection ${\sf pr}{\overline C^{\,\prime}}
\overset{\text{def}}{=}{\overline C^{\,\prime}} \cap S^{n-1} \subset
{\sf pr}C\overset{\text{def}}{=}C \cap S^{n-1}$, where $S^{n-1}$ is the unit sphere
in $\oR^n$. Being given a cone $C$ in $y$-space, we associate with $C$ a closed convex
cone $C^*$ in $\xi$-space which is the set $C^*=\bigl\{\xi \in \oR^n \mid \langle \xi,y
\rangle \geq 0,\forall\,\,y \in C \bigr\}$. The cone $C^*$ is called the {\em dual cone}
of $C$. In the sequel, it will be sufficient to assume for our purposes that the open
connected cone $C$ in $\oR^n$ is an open convex cone with vertex at the origin
and proper. By $T(C)$ we will denote the set $\oR^n+iC \subset \oC^n$.
If $C$ is open and connected, $T(C)$ is called the tubular radial domain in
$\oC^n$, while if $C$ is only open $T(C)$ is referred to as a tubular cone. In the
former case we say that $f(z)$ has a boundary value $U=BV(f(z))$ in ${\mathfrak H}^\prime$
as $y \rightarrow 0$, $y \in C$ or $y \in C^\prime \Subset C$, respectively, if
for all $\psi \in {\mathfrak H}$ the limit
\[
\langle U, \psi \rangle=\lim_{{\substack{y \rightarrow 0 \\
y \in C~{\rm or}~C^\prime}}} \int_{\oR^n} f(x+iy)\psi(x) d^nx\,\,,
\]
exists. We will deal with tubes defined as the set of all points $z \in \oC^n$
such that
\[
T(C)=\Bigl\{x+iy \in \oC^n \mid
x \in \oR^n, y \in C, |y| < \delta \Bigr\}\,\,,
\]
where $\delta > 0$ is an arbitrary number.

An important example of tubular radial domain used in quantum field theory
is the forward light-cone
\[
V_+=\Bigl\{z \in \oC^n \mid {\rm Im}\,z_1 >
\Bigl(\sum_{i=2}^n {\rm Im}^2\,z_i \Bigr)^{\frac{1}{2}}, {\rm Im}\,z_1 > 0 \Bigr\}\,\,.  
\]

Let $C$ be an open convex cone, and let $C^\prime \Subset C$.
Let $B[0;r]$ denote a {\bf closed} ball of the
origin in $\oR^n$ of radius $r$, where $r$ is an arbitrary positive
real number. Denote $T(C^\prime;r)=\oR^n+i\bigl(C^\prime \setminus
\bigl(C^\prime \cap B[0;r]\bigr)\bigr)$.
We are going to introduce a space of holomorphic functions
which satisfy certain estimate according to Carmichael~\cite{Carmi1}.
We want to consider the space consisting of holomorphic functions $f(z)$
such that
\begin{equation}
\bigl|f(z)\bigr|\leq {\boldsymbol{\sf M}}(C^\prime)(1+|z|)^N e^{h_{C^*}(y)}
\,\,,\quad z \in T(C^\prime;r)\,\,,
\label{eq31} 
\end{equation}
where $h_{C^*}(y)=\sup_{\xi \in C^*}\langle \xi,y \rangle$ is the supporting
function of $C^*$, ${\boldsymbol{\sf M}}(C^\prime)$ is a constant that depends
on an arbitrary compact cone $C^\prime$ and $N$ is a non-negative real number.
The set of all functions $f(z)$ which are holomorphic in $T(C^\prime;r)$ and
satisfy the estimate (\ref{eq31}) will be denoted by $\boldsymbol{{\mathscr H}^o_c}$.

\begin{remark}
The space of functions $\boldsymbol{{\mathscr H}^o_c}$ constitutes a generalization
of the space ${\mathfrak A}_{_\omega}^i$ of Sebati\~ao e Silva~\cite{Tiao1} and the
space $\mona_{_\omega}$ of Hasumi~\cite{Hasumi} to arbitrary tubular radial domains
in $\oC^n$.
\end{remark}

\begin{lemma}[\cite{Carmi1,DanHenri}]
Let $C$ be an open convex cone, and let $C^\prime \Subset C$.
Let $h(\xi)=e^{k|\xi|}g(\xi)$, $\xi \in \oR^n$, be
a function with support in $C^*$, where $g(\xi)$ is a bounded continuous
function on $\oR^n$. Let $y$ be an arbitrary but fixed point of
$C^\prime \setminus \bigl(C^\prime \cap B[0;r]\bigr)$. Then
$e^{-\langle \xi,y \rangle}h(\xi) \in L^2$, as a function of $\xi \in \oR^n$.
\label{lemma0}
\end{lemma}

\begin{definition}
We denote by $H^\prime_{C^*}(\oR^n;O)$ the subspace of $H^\prime(\oR^n;O)$
of distributions of exponential growth with support in the cone $C^*$:
\begin{equation}
H^\prime_{C^*}(\oR^n;O)=\Bigl\{V \in H^\prime(\oR^n;O) \mid
\supp(V) \subseteq C^* \Bigr\}\,\,. 
\label{eq31'} 
\end{equation}
\label{Def1}
\end{definition}

\begin{lemma}[\cite{Carmi1,DanHenri}]
Let $C$ be an open convex cone, and let $C^\prime \Subset C$.
Let $V=D^\gamma_\xi[e^{h_K(\xi)}g(\xi)]$, where
$g(\xi)$ is a bounded continuous function on $\oR^n$ and $h_K(\xi)=k|\xi|$
for a convex compact set $K=\bigl[-k,k\bigr]^n$. Let
$V \in H^\prime_{C^*}(\oR^n;O)$. Then $f(z)=(2\pi)^{-n}
\bigl\langle V,e^{-i\langle \xi,z \rangle}\bigr\rangle$ is an element of
$\boldsymbol{{\mathscr H}^o_c}$.
\label{lemma1}
\end{lemma}

We now shall define the main space of holomorphic functions with which this paper
is concerned. Let $C$ be a proper open convex cone, and let $C^\prime \Subset C$.
Let $B(0;r)$ denote an {\bf open} ball of the origin in $\oR^n$ of radius
$r$, where $r$ is an arbitrary positive real number. Denote $T(C^\prime;r)=
\oR^n+i\bigl(C^\prime \setminus \bigl(C^\prime \cap B(0;r)\bigr)\bigr)$. Throughout
this section, we consider functions $f(z)$ which are holomorphic in
$T(C^\prime)=\oR^n+iC^\prime$ and which satisfy the estimate (\ref{eq31}),
with $B[0;r]$ replaced by $B(0;r)$. We denote this space by
$\boldsymbol{{\mathscr H}^{*\,o}_c}$. We note that $\boldsymbol{{\mathscr H}^{*\,o}_c}
\subset \boldsymbol{{\mathscr H}^{o}_c}$ for any open convex cone $C$. Put
${\mathscr U}_c=\boldsymbol{{\mathscr H}^{*\,o}_c}/\boldsymbol{\Pi}$, that is,
${\mathscr U}_c$ is the quotient space of $\boldsymbol{{\mathscr H}^{*\,o}_c}$
by set of pseudo-polynomials $\boldsymbol{\Pi}$.

\begin{definition}
The set ${\mathscr U}_c$ is the space of tempered ultrahyperfunctions corresponding
to a proper open convex cone $C \subset \oR^n$.
\end{definition}

A useful property of tempered ultrahyperfunctions corresponding to a proper cone
is the distributional boundary value theorem concerning analytic functions.
The following theorem shows that functions in
$\boldsymbol{{\mathscr H}^{*\,o}_c}$ have distributional boundary values in
${\mathfrak H}^\prime$. Further, it shows that functions in
$\boldsymbol{{\mathscr H}^{*\,o}_c}$ satisfy a strong boundedness property in
${\mathfrak H}^\prime$.

\begin{theorem}[\cite{Daniel1}]
Let $C$ be an open convex cone, and let $C^\prime \Subset C$.
Let $V=D^\gamma_\xi[e^{h_K(\xi)}g(\xi)]$, where
$g(\xi)$ is a bounded continuous function on $\oR^n$ and $h_K(\xi)=k|\xi|$
for a convex compact set $K=\bigl[-k,k\bigr]^n$. Let
$V \in H^\prime_{C^*}(\oR^n;\oR^n)$. Then

\,\,\,$(i)\quad f(z)=(2\pi)^{-n}\bigl\langle V,e^{-i\langle \xi,z \rangle}\bigr\rangle$
is an element of $\boldsymbol{{\mathscr H}^{*\,o}_c}$,

\,\,\,$(ii)\quad \bigl\{f(z) \mid y={\rm Im}\,z \in C^\prime \Subset C, |y| \leq Q\bigr\}$
is a strongly bounded set in ${\mathfrak H}^\prime$, where $Q$ is an arbitrarily but fixed
positive real number,

\,\,\,$(iii)\quad f(z) \rightarrow {\mathscr F}^{-1}[V] \in {\mathfrak H}^\prime$ in the
strong (and weak) topology of ${\mathfrak H}^\prime$ as $y={\rm Im}\,z \rightarrow 0$,
$y \in C^\prime \Subset C$.
\label{theorem1}
\end{theorem}

The functions $f(z) \in \boldsymbol{{\mathscr H}^{*\,o}_c}$ can be recovered as
the (inverse) Fourier-Laplace transform\footnote{The convention of signs in the
Fourier transform which is used here one leads us to consider the inverse
Fourier-Laplace transform.} of the constructed distribution $V \in
H^\prime_{C^*}(\oR^n;\oR^n)$. This result is a generalization of the Paley-Wiener-Schwartz
theorem for the setting of tempered ultrahyperfunctions.

\begin{theorem}[\cite{Daniel1}]
Let $f(z) \in \boldsymbol{{\mathscr H}^{*\,o}_c}$, where $C$ is an open convex cone.
Then the distribution $V \in H^\prime_{C^*}(\oR^n;O)$ has a uniquely
determined inverse Fo\-uri\-er-Laplace transform $f(z)=(2\pi)^{-n}
\bigl\langle V,e^{-i\langle \xi,z \rangle} \bigr\rangle$ which is holomorphic in
$T(C^\prime)$ and satisfies the estimate (\ref{eq31}), with $B[0;r]$ replaced by
$B(0;r)$.
\label{PWSTheo} 
\end{theorem}

We finish this section with two results proved in Ref.~\cite{Daniel1}, which
will be used in the applications of Section \ref{Section4}.

\begin{theorem}[Tempered ultrahyperfunction version of edge of the wedge
theorem]
Let $C$ be an open cone of the form $C=C_1 \cup C_2$, where each $C_j$, $j=1,2$,
is a proper open convex cone. Denote by $\boldsymbol{ch}(C)$ the convex hull of the
cone $C$. Assume that the distributional boundary values of two holomorphic functions
$f_j(z) \in \boldsymbol{{\mathscr H}^{*\,o}_{c_j}}$ $(j=1,2)$ agree, that
is, $U=BV(f_1(z))=BV(f_2(z))$, where $U \in {\mathfrak H}^\prime$ in accordance with
the Theorem \ref{theorem1}. Then there exists $F(z) \in
\boldsymbol{{\mathscr H}^{o}_{{\boldsymbol{ch}(C)}}}$
such that $F(z)=f_j(z)$ on the domain of definition of each $f_j(z)$, $j=1,2$.
\label{EWTheo}
\end{theorem}

\begin{theorem}
Let $C$ be some open convex cone. Let $f(z) \in \boldsymbol{{\mathscr H}^{*\,o}_{c}}$.
If the boundary value $BV(f(z))$ of $f(z)$ in the sense of tempered ultrahyperfunctions
vanishes, then the function $f(z)$ itself vanishes.
\label{UniTheo}
\end{theorem}

\section{Reeh-Schlieder-Type Theorem for Ultrahyperfunctional Quantum Fields}
\label{Section4}

\begin{definition}
Assume we are given a Hibert space $\mathscr H$. According
to~\cite[Proposition 4.1]{BruNa1}, we define the space of $\mathscr H$
valued tempered ultrahyperfunctions to be the set of all continuous linear
mapping from ${\mathfrak H}(T(\oR^{4m})$ to $\mathscr H$.
\end{definition}

\begin{theorem}
For any non-empty open set $X \subset \oR^4$, the set of vectors
of the form $\Phi(f_1) \cdots \Phi(f_m) \Omega_o$
with $f_j(x) \in {\mathfrak H}(T(\oR^4))$ and $x={\rm Re}\,z \in X$, is dense in
$\mathscr H$.
\label{RSTheo}  
\end{theorem}

\begin{proof} Denote by $D_o$ the minimal common invariant domain, which is
assumed to be dense, of the field operators in the Hilbert space ${\mathscr H}$ of states,
{\em i.e.}, the vector subspace of ${\mathscr H}$ that is spanned by the vacuum state
$\Omega_o$ and by the set of vectors
\[
\bigl\{\Phi(f_1) \cdots \Phi(f_m) \Omega_o \mid \supp\,f_j(x)
\subset X, m \in \oN \bigr\}. 
\]
Let $\Psi \in {\mathscr H}$ be orthogonal to all vectors of the form
$\Phi(f_1) \cdots \Phi(f_m) \Omega_o \in D_o$. Then, it is required to prove that
$\Psi$ is identically zero.

According to Ref.~\cite{BruNa1},
\[
\Bigl[{\mathfrak H}(T(\oR^4))\Bigr]^m \ni (f_1,\dots,f_m)\rightarrow
\langle \Psi,\Phi(f_1) \cdots \Phi(f_m) \Omega_o\rangle 
\]
is a multilinear functional in each $f_j \in {\mathfrak H}(T(\oR^4))$ separately 
with all the others $f_i \in {\mathfrak H}(T(\oR^4))$, $i \not= j$, kept fixed.
However, then the Theorem \ref{KernelTheo} implies that the functional
$\langle \Psi,\Phi(f_1) \cdots \Phi(f_m) \Omega_o\rangle$ has
a uniquely determined extension to a tempered ultrahyperfunction 
$\boldsymbol{F}_\Psi \in {\mathscr U}_c(\oR^{4m})$ such that 
\begin{equation}
\boldsymbol{F}_\Psi(f^{(m)})=\int d^4z_1 \cdots d^4z_m\,\,
{\mathfrak F}^{(1)}_\Psi(z_1,\ldots,z_m) f^{(m)}(x_1,\ldots,x_m)\,\,,
\label{gfunction1}
\end{equation}
for every $\Psi \in {\mathscr H}$, where ${\mathfrak F}^{(1)}_\Psi(z_1,\ldots,z_m)=
\langle \Psi,\Phi(z_1) \cdots \Phi(z_m) \Omega_o \rangle$. According to the
arguments of Section IV.C of Ref.~\cite{BruNa1}, the Fourier transform
$\widehat{\boldsymbol{F}_\Psi}$ vanishes unless each four-momentum
variable lies in the physical spectrum. Hence, we can apply Theorem \ref{PWSTheo}
to conclude that $\boldsymbol{F}_\Psi$ is holomorphic in the set
$T(V^\prime_+;r)=\oR^{4m}+i\bigl(V^\prime_+ \setminus
\bigl(V^\prime_+ \cap B[0;r]\bigr)\bigr)$, with $V^\prime_+ \Subset V_+$.
Then, by Theorem \ref{theorem1}, we have that $\boldsymbol{F}_\Psi|_{_X}$
is the boundary value of ${\mathfrak F}^{(1)}_\Psi$ when $V^\prime_+ \ni y_1
\rightarrow 0$, $V^\prime_+ \ni (y_j-y_{j-1}) \rightarrow 0$, $j=2,\ldots,m$.
Furthermore the function ${\mathfrak F}^{(2)}_\Psi(z_1,\ldots,z_m)=
\overline{{\mathfrak F}^{(1)}_\Psi(\bar{z}_1,\ldots,\bar{z}_m)}$ is holomorphic
in the set $T(V^\prime_-;r)=\oR^{4m}+i\bigl(V^\prime_- \setminus
\bigl(V^\prime_- \cap B[0;r]\bigr)\bigr)$, with $V^\prime_-= - V^\prime_+$ and
$\boldsymbol{F}_\Psi|_{_X}$ is the boundary value of ${\mathfrak F}^{(2)}_\Psi$
when $V^\prime_- \ni y_1 \rightarrow 0$, $V^\prime_- \ni (y_j-y_{j-1})
\rightarrow 0$, $j=2,\ldots,m$.
By hypothesis, $\boldsymbol{F}_\Psi|_{_X}$ vanishes on a non-empty open real set
$x_1,\ldots,x_m \in X^{m}$, since $D_o$ spans the Hibert space $\mathscr H$.
Therefore we can apply the Edge of the Wedge Theorem \ref{EWTheo} in order to
show that ${\mathfrak F}^{(1)}_\Psi$ and ${\mathfrak F}^{(2)}_\Psi$ have a
common analytic continuation ${\mathfrak F}_\Psi$. Since ${\mathfrak F}_\Psi$
vanishes on $X^{m}$, it vanishes together with ${\mathfrak F}^{(1)}_\Psi$
identically by Theorem \ref{UniTheo}. This shows that $\Psi$ is even
orthogonal to the set $\bigl\{\Phi(f_1) \cdots \Phi(f_m) \Omega_o \mid
f_j(x) \in {\mathfrak H}(T(\oR^4)), j=1,\ldots,m \bigr\}$.
We conclude that $\Psi \in D_o^\perp=\{0\}$. This completes the proof of theorem.   
\end{proof}

In what follows, we give Reeh-Schlieder theorem for NCQFT in the setting of
tempered ultrahyperfunctions.
When referring to NCQFT one should have in mind the deformation of the
ordinary product of fields. In terms of complex variables, this deformation is
performed through the star product extended for noncoinciding points via the
functorial relation
\begin{align}
\varphi(z_1)\star \cdots \star \varphi(z_n)=
\prod\limits_{i<j}
\exp \left({\frac{1}{2}\theta^{\mu\nu}\frac{\partial}{\partial z_i^\mu}
\wedge \frac{\partial}{\partial \bar{z}_j^\nu}} \right)
\varphi(z_1)\cdots \varphi(z_n)\,\,.
\label{Moyalprod}
\end{align}
For coinciding points $z_1=z_2=\cdots=z_n$ the product (\ref{Moyalprod}) becomes
identical to the multiple Moyal $\star$-product. We consider NCQFT in the sense of
a field theory on a non-commutative space-time encoded by a Moyal product. 

In this point, a few comments about the NCQFT are in order.
Generalizing the Wightman axioms to NCQFT is not as simple, especially the
Poincar\'e symmetry. It is well known that due to the constant matrix
$\theta$, the Poincar\'e symmetry is not preserved in NCQFT. Furthemore,
the existence of hard infrared singularities in the non-planar sector of
the theory can destroy the {\em tempered} nature of the Wightman functions.
And more, how can the local commutativity condition be described in a
field theory with a fundamental length? The analysis in Ref.~\cite{DZH} has shown
that the sequence of vacuum expectation values of a NCQFT in terms of tempered
ultrahyperfunctions satisfies a number of specific properties, which actually characterize
a NCQFT in terms of tempered ultrahyperfunctions. We summarize these below (for details
see~\cite{DZH}):

\newcounter{numero}
\setcounter{numero}{0}
\def\Prop{\addtocounter{numero}{1}\item[{$\boldsymbol{\sf P_{\thenumero}}$}]}
\begin{enumerate}

\Prop ${\mathfrak W^\star_0}=1$, ${\mathfrak W^\star_m} \in {\mathscr U}_c(\oR^{4m})$
for $n \geq 1$, and ${\mathfrak W^\star_m}(f^*)=\overline{{\mathfrak W^\star_m}(f)}$,
for all $f \in {\mathfrak H}(T(\oR^{4m}))$, where 
${\mathfrak W}_m^{\star}(z_{1},\ldots,z_{m})\overset{\text{\rm def}}
{=}\langle\Omega_o \mid \Phi(z_1)\star \cdots \star \Phi(z_m) \mid \Omega_o\rangle$
and $f^*(z_1,\ldots,z_m)=\overline{f(\bar{z}_1,\ldots,\bar{z}_m)}$.

\bigskip

\Prop The Wightman functionals ${\mathfrak W^\star_m}$ are invariant under
the {\em twisted} Poincar\'e group

\bigskip

\Prop Spectral condition. Since the Fourier transformation of tempered
ultrahyperfunctions are distributions, the spectral condition is not
so much different from that of Schwartz distributions. Thus, for every
$m \in \oN$, there is $\widehat{{\mathfrak W}}^\star_{m} \in
H_{V^*}^\prime(\oR^{4m},\oR^{4m})$~\cite{BruNa1}, where
\begin{equation}
H^\prime_{V^*}(\oR^{4m},\oR^{4m})=\Bigl\{V \in H^\prime(\oR^{4m},\oR^{4m}) \mid
\supp\,(\widehat{{\mathfrak W}}^\star_{m}) \subset V^* \Bigr\}\,\,, 
\label{EQ31'} 
\end{equation}
with $V^*$ being the properly convex cone defined by
\[
\Bigm\{(p_1,\ldots,p_m)\in {\Bbb R}^{4m}\,\,\bigm|\,\,\sum_{j=1}^{m}p_j=0,\,\,
\sum_{j=1}^{k}p_j \in {\overline V}_+,\,\,k=1,\dots,m-1 \Bigm\}\,\,,
\]
where ${\overline V}_+=\{(p^0,{\boldsymbol p}) \in \oR^4 \mid p^2 \geq 0,
p^0 \geq 0\}$ is the closed forward light cone.

\bigskip

\Prop Extended local commutativity condition.

\bigskip
 
\Prop For any finite set $f_o,f_1,\ldots,f_N$ of test functions such that
$f_o \in \oC$, $f_j \in {\mathfrak H}(T(\oR^{4j}))$ for $1 \leq j \leq N$,
one has
\begin{align*}
\sum_{k,\ell=0}^N {\mathfrak W}^\star_{k+\ell}(f_k^* \otimes f_\ell)
\geq 0\,\,.
\end{align*}

\end{enumerate} 
\begin{remark}
The tempered ultrahyperfunctions ${\mathfrak W}^\star_m \in {\mathscr U}_c(\oR^{4m})$
have been called {\em non-commu\-ta\-tive} Wightman functions in~\cite{DZH}.
\end{remark}

\begin{theorem}[Reeh-Schlieder Theorem for NCQFT]
Suppose that the hypotheses of Theorem \ref{RSTheo} hold except
that instead of vectors of the form $\Phi(f_1) \cdots \Phi(f_m) \Omega_o$,
we have vectors of the form $\Phi(f_1)\star \cdots \star \Phi(f_m) \Omega_o$. Then
the conclusions of Theorem \ref{RSTheo} again hold.
\end{theorem}

\begin{proof}
For this purpose, we consider the functional
\[
\langle \Psi,\Phi(f_1)\star \cdots \star \Phi(f_m) \Omega_o \rangle=
\prod\limits_{i<j}
\exp \left({\frac{1}{2}\theta^{\mu\nu}\frac{\partial}{\partial z_i^\mu}
\wedge \frac{\partial}{\partial \bar{z}_j^\nu}} \right)
\langle \Psi,\Phi(f_1)\cdots\Phi(f_m) \Omega_o \rangle
\]
One first notes that the formula above simplifies considerably the
proof of theorem in the case of NCQFT in terms of tempered ultrahyperfunctions,
since $\langle \Psi,\Phi(f_1)\cdots\Phi(f_m) \Omega_o \rangle$ is representable
by means of holomorphic functions (the holomorphy properties of the functions
under consideration are discussed in Ref.~\cite{DZH}). Thus the star product
coincides with the regular product of fields
\begin{equation}
\langle \Psi,\Phi(f_1)\star \cdots \star \Phi(f_m) \Omega_o \rangle=
\langle \Psi,\Phi(f_1)\cdots\Phi(f_m) \Omega_o \rangle\,\,.
\label{gfunction}
\end{equation}
This means that a NCQFT in terms of tempered ultrahyperfunctions is unchanged
by the deformation of the product. Therefore, the conclusions of Theorem \ref{RSTheo}
again hold.
\end{proof}

\begin{remark}
In~\cite{Vernov} the Wightman functions were written as follows:
\[
{\mathfrak W}_m^{\tilde\star}(z_{1},\ldots,z_{m})\overset{\text{\rm def}}
{=}\langle\Omega_o \mid \Phi(z_1)\tilde\star \cdots \tilde\star\, \Phi(z_m)
\mid \Omega_o\rangle\,\,,
\]
where the meaning of $\tilde\star$ depends on the considered case. In particular,
if $\tilde\star = 1$, we obtain the standard form ${\mathfrak W}_m(z_{1},\ldots,z_{m})=
\langle\Omega_o \mid \Phi(z_1) \cdots \Phi(z_m) \mid \Omega_o\rangle$ adopted in~\cite{AGVM}.
On the other hand, if $\tilde\star = \star$, this choice corresponds to the Wightman functions
introduced in~\cite{Chai1}. In this case, the non-commutativity is manifested not only at
coincident points but also in their neighborhood. The Equation (\ref{gfunction}) reflects
the fact that the axiomatic approach to the NCQFT in terms of tempered ultrahyperfunctions is
{\it independent} of the concrete type of the $\tilde\star$-product (similar conclusion was
obtained in~\cite{Vernov}).
\end{remark}

\section{Final Considerations}
\label{Section5}
In the present paper, we to consider a quantum field theory on
non-commuta\-ti\-ve space-times in terms of the tempered ultrahyperfunctions of
Sebasti\~ao e Silva corresponding to a convex cone, within the framework formulated
by Wightman. Tempered ultrahyperfunctions are representable by means of holomorphic
functions. As is well known there are certain advantages to be gained from the
representation of distributions in terms of holomorphic functions. In
particular, for non-commutative theories the product of fields involving
the $\star$-product has the same form as the ordinary product of fields (effects
of non-commutativity are nontrivial in the formula with real variables).
In light of this result, we show that the Reeh-Schlieder property,
proved in the framework of local QFT, also holds for states of quantum fields
on non-commutative space-times.

\section*{Acknowledgments}
The author would like to express his gratitude to Afr\^anio R. Pereira,
Winder A.M. Melo and to the Departament of Physics of the Universidade Federal
de Vi\c cosa (UFV) for the opportunity of serving as Visiting Researcher.


\end{document}